# Single-electron charge transfer into putative Majorana and trivial modes in individual vortices

*Jian-Feng Ge,[1] Koen M. Bastiaans,[1, 2] Damianos Chatzopoulos,[1] Doohee Cho,[3] Willem O. Tromp,[1] Tjerk Benschop,[1] Jiasen Niu,[1] Genda Gu,[4] Milan P. Allan[1]\**

[1] *Leiden Institute of Physics, Leiden University, 2333 CA Leiden, The Netherlands*
[2] *Department of Quantum Nanoscience, Kavli Institute of Nanoscience, Delft University of Technology, 2628 CJ Delft, The Netherlands*
[3] *Department of Physics, Yonsei University, Seoul 03722, Republic of Korea*
[4] *Condensed Matter Physics and Materials Science Department, Brookhaven National Laboratory, Upton, NY, 11973, USA*

*Corresponding author. Email: allan@physics.leidenuniv.nl

## ABSTRACT

**Majorana bound states are putative collective excitations in solids that exhibit the self-conjugate property of Majorana fermions—they are their own antiparticles. In iron-based superconductors, zero-energy states in vortices have been reported as potential Majorana bound states, but the evidence remains controversial. Here, we use scanning tunneling noise spectroscopy to study the tunneling process into vortex bound states in the conventional superconductor $NbSe_2$, and in the putative Majorana platform $FeTe_{0.55}Se_{0.45}$. We find that tunneling into vortex bound states in both cases exhibits charge transfer of a single electron charge. Our data for the zero-energy bound states in $FeTe_{0.55}Se_{0.45}$ exclude the possibility of Yu–Shiba–Rusinov states and are consistent with both Majorana bound states and trivial vortex bound states. Our results open an avenue for investigating the exotic states in vortex cores and for future Majorana devices, although further theoretical investigations involving charge dynamics and superconducting tips are necessary.**

## INTRODUCTION

When a type-II superconductor is exposed to magnetic fields, vortices emerge as line defects where the order parameter vanishes, quantized magnetic flux penetrates the superconductor, and localized low-energy bound states form in the vortex cores. The nature of the vortex bound state is mysterious in many unconventional superconductors. Recently, much focus is on iron-based superconductors, where topologically nontrivial superconductivity and elusive Majorana bound states have been predicted to exist in vortex cores[1].

So far, the most often reported signature for Majorana bound states is a peak in tunneling differential conductance at zero bias voltage. This signature is readily accessible by experiments, but it is not conclusive proof of the Majorana character of a state[2–5]. Other topologically trivial bound states, including Yu–Shiba–Rusinov (YSR) states, can also show the same zero-bias conductance peak, as demonstrated in proximitized superconducting nanowires[6,7]. Further, Caroli–de Gennes–Matricon (CdGM) states in the vortex cores are difficult to differentiate from Majorana bound states, because the former could also appear at zero energy[4, 8-12].

Zero-bias conductance peaks in full-flux-quantum vortex cores are also the main evidence for Majorana bound states in the iron-based superconductor FeTe$_{0.55}$Se$_{0.45}$[13], which is the focus of this study. However, controversy remains as the absence of zero-energy bound states has been reported[14,15]. It is still being debated whether the additional observation[16] — a saturating conductance at roughly two-thirds of the expected quantized value $2e^2/h$—is a strong argument for the Majorana character ($h$ is the Planck constant and $e$ is the elementary charge). The issue is that such saturating behavior at an arbitrary conductance near $2e^2/h$ has been observed for YSR states[17] as well; these states are present in FeTe$_{0.55}$Se$_{0.45}$ and may also appear as a conductance peak at zero bias[18]. Furthermore, it was pointed out that the simple approximations of the Fu–Kane model are not likely applicable to the system of vortices in iron-based superconductors[19], which brings the exact nature of the zero-energy vortex bound states in FeTe$_{0.55}$Se$_{0.45}$ into question.

New local probes are thus desired to investigate the electronic properties of the vortex bound states in iron-based superconductors. It has been widely investigated theoretically how shot noise could act as a tell-tale probe to distinguish between trivial and Majorana bound states in vortex matter and nanowires[20–36], but experiments have not been possible. The principle behind most theoretical proposals is that Majorana bound states induce resonant Andreev reflection: an incident electron from the coupling lead, when tunneling into a Majorana bound state, is reflected as a hole with unity probability[22]. Such a resonant Andreev process is predicted to generate unique Majorana signatures that are absent for trivial fermionic states.

These theoretical studies form the motivation for the shot noise measurements on individual vortex cores that we present here. We measure both the vortex bound states in a conventional superconductor NbSe$_2$ and the putative Majorana bound states in vortices of FeTe$_{0.55}$Se$_{0.45}$ at temperature $T = 2.3$ K. While we argue that our results do not represent a smoking gun experiment for the existence of Majorana bound states, they allow us to exclude YSR states as the origin of the zero-bias conductance peak, and they give an experimental insight into these bound states.

Shot noise is, at its core, a consequence of the discreteness of charge. Because of this, tunneling is a Poissonian process, and the noise spectral density $S$ is proportional to the time-averaged current $I$,

$$S = 2q^*|I|. \tag{1}$$

Shot noise thus allows probing two quantities that are not visible in the time-averaged current: the effective charge $q^*$ of the charge carriers and possible correlations between them in electronic matters[37]. The former has been used to measure fractional charges in mesoscopic quantum hall systems[38], and the latter has been used to measure the vanishing noise at the quantum conductance of break junctions[39].

Despite a large number of theoretical studies on shot noise tunneling into vortex cores, there has been no experiment so far. The challenge is that one needs high enough sensitivity to measure the change in $q^*$ from noise, with nanometer resolution to locate individual vortices. This nanoscale resolution is not feasible in mesoscopic setups where noise measurements have been widely applied.

Recently, we have developed scanning tunneling noise microscopy (STNM), which combines scanning tunneling microscopy (STM) and noise spectroscopy, allowing us to measure the effective tunneling charge with atomic resolution[40]. To do so, we build a cryogenic megahertz amplifier that works in parallel with the usual dc measurements, as illustrated in Fig. 1a. STNM has revealed paired electrons in superconductors[41,42]. STNM also allows to measure shot noise exactly at the core of an individual vortex, which provides a direct and local extraction of the effective charge of the tunneling process into vortex bound states.

Here, we measure two different materials: the iron-based superconductor FeTe$_{0.55}$Se$_{0.45}$, which is conjectured to host putative Majorana bound states, and the conventional superconductor 2H-NbSe$_2$ as a comparison. We use a tip with an apex made out of Pb, which is a type-I, $s$-wave superconductor with a relatively large gap $\Delta_t \sim 1.3$ meV. We choose to use a superconducting tip in this study for two reasons: first, a superconducting tip provides a superior energy resolution without the limitation from thermal broadening as in the case of a normal-metal tip, i.e. ~0.25 meV (See Supplementary Note 1) instead of ~$3.5k_BT = 0.70$ meV, where $k_B$ is the Boltzman constant; second, as a consequence of a convolution with the density of states of the superconducting tip, the tunneling signal into a zero-energy vortex bound state is effectively shifted from the Fermi level to $\pm\Delta_t$ (illustrated in Supplementary Fig. 2)[6]. This shift circumvents the challenge of measuring shot noise at zero bias voltage.

**RESULTS**

**Vortex bound states in NbSe$_2$ and FeTe$_{0.55}$Se$_{0.45}$**

We first image the subgap electronic structure of the vortices in NbSe$_2$. We introduce vortices by applying an external magnetic field $B = 0.1$ T perpendicular to the sample surface (the upper critical field of the tip $B_c \sim 0.7$ T, see Supplementary Fig. 1). Because vortices have the strongest enhancement in density of states at the Fermi level of the sample, they are visible as enhanced differential conductance at the energy $|E| = \Delta_t$ when using a superconducting tip (Supplementary Fig. 2a). Figure 1b shows a spatially resolved image of the differential conductance taken with a sample bias $V_{bias} = -\Delta_t/e$, revealing the full-flux-quantum ($h/2e$) vortex lattice, with each vortex in the characteristic sixfold star shape for NbSe$_2$[43,44]. We then take differential conductance maps $g(E, \mathbf{r})$ on a fine spatial grid around an individual vortex as shown in Fig. 2. Away from the vortex core, the spectrum in Fig. 2b shows an energy gap with a size of $2(\Delta_t + \Delta_s)$, where $\Delta_s = 1.0$ meV is the superconducting gap of the sample. On the other hand, the spectrum measured at the core center develops two peaks at $\pm\Delta_t$, which translates to a zero-bias conductance peak for a spectrum taken with a normal-metal tip[6]. This translation is confirmed by a deconvolution procedure[18,45] that extracts the local density of states of the sample (see Supplementary Note 2); as expected, the resulting density of states has a peak at zero energy (Fig. 2e). Deconvolution of spectra along a linecut through the vortex reveals that the zero-bias peak splits away from the core into two dispersing peaks, which eventually merge to the gap edges. These dispersing states are consistent with previous studies[43] of NbSe$_2$ and the expectations of many closely-spaced (on the order of 40 µeV) CdGM bound states from solving the Bogoliubov–de Gennes equations[46], where the peak at a longer distance from the core center corresponds to a CdGM bound state with a larger angular momentum.

In contrast to the dispersing CdGM bound states in vortex cores of NbSe$_2$, a Majorana bound state is topologically protected such that its energy is locked at the Fermi level[1]. This is exactly what we observe, in agreement with the literature[13,16], in tunneling differential conductance measurements on vortices of FeTe$_{0.55}$Se$_{0.45}$ (Fig. 3). The zero-bias conductance peak does not split (Fig. 3c, f) as in the case of NbSe$_2$ (See Supplementary Note 7); instead, the non-split bound state extends ~8 nm spatially across the vortex core (Supplementary Fig. 2), identical to the states observed and interpreted as Majorana bound states in Refs. [13,15,16]. In principle, one expects a pair of peaks at $\pm\Delta_t$ with an equal amplitude in the differential conductance spectrum when tunneling into a Majorana state[47]. We find, that the pairs of peaks $\pm\Delta_t$ we observe at every spectrum on vortex are asymmetric (e.g., the red spectrum in Fig. 3b), which may indicate, the presence of accompanying states such as CdGM or YSR states, assuming a superconducting tip with a particle-hole symmetric density of states. We also note that the hybridization between Majorana bound states in a vortex lattice could also split the conductance peaks owing to the spatial overlap of Majorana wavefunctions. However, since the average distance between vortices in Fig. 1c is about 120 nm, the energy splitting for the putative Majorana bound states is on the order of 1 μeV[48].

Before discussing shot-noise, we comment on what conventional conductance spectroscopy can contribute to distinguishing Majorana, CdGM, or YSR states. CdGM states are expected to be at finite energy instead of zero energy, but the energy difference can be small, and additional effects might shift the energy[8]. CdGM states have been observed in FeTe$_{0.55}$Se$_{0.45}$—surprisingly only in a subset of vortices[13,14]. In these vortices, the lowest energy levels have been reported as small as ~0.1 meV. From Lorentzian fits (Supplementary Fig. 2), our results show that the energy of the zero-bias peak is $0 \pm 50$ μeV, much smaller than the energy of the lowest-lying CdGM bound states reported. Furthermore, a previous high-energy-resolution study[15] has shown that while some vortices show non-zero-bias peaks, associated with CdGM states, the majority (~80% at $B = 1.0$ T) show zero-bias peaks having an energy of $0 \pm 20$ μeV, which is evidence to exclude CdGM states. Based on these studies and the statistics therein, combined with our measured electronic structure, we deduce that the probability that all the vortices measured here only have CdGM states is less than 0.8%. Therefore, the zero-energy state we observe here is in agreement with the putative Majorana bound state previously reported – with the caveat that it has recently been shown that a CdGM state can imitate a zero-energy state[10,12]. We end this discussion by noting that the possibility of trivial YSR states, which can exist at zero bias[18], has been investigated much less.

**Effective charge inside and outside of vortex cores**

The key advance of this study is high-sensitivity, atomic-scale noise spectroscopy that allows to extract the effective charge $q^*$ transferred when tunneling into vortex bound states. In the tunneling regime where the transparency is small, we parametrize any changes in noise, including the so-called Fano factor, via the effective charge $q^*$ in Eq. (1) (see Methods for different definitions of the Fano factor). For example, in the simplest case of electron tunneling, the transferred charge of each tunneling event is a single electron charge ($q^* = 1e$), as expected from Poissonian statistics. In contrast, when Andreev reflection takes place such that an incident electron is reflected as a hole, two electron charges ($q^* = 2e$) are effectively transferred per event. Since our measurements are performed at a finite temperature $T$, the current noise in the tunnel junction with resistance $R_J$ consists of shot noise and thermal current noise $4k_BT/R_J$, and takes the form of[37]

$$S = 2q^*(V_{bias}/R_J)\coth(q^*V_{bias}/2k_BT). \tag{2}$$

This equation, which reduces to Eq. (1) at zero temperature, allows us to extract the effective charge $q^*$ as a function of bias voltage. Note that we keep $R_J$ constant during noise spectroscopy by the changing tip-sample distance in a slow feedback loop (See Methods and Supplementary Note 6 for details).

We start by measuring current noise at $B = 0.1$ T at locations far away from the vortex cores. There, one expects the noise to correspond to an effective charge of $q^* = 1e$ at bias energies larger than the gap, $|eV_{bias}| > (\Delta_t + \Delta_s)$. At these energies, the tunneling of Bogoliubov quasiparticles dominates the noise. Around the gap energy $\pm(\Delta_t + \Delta_s)$, one then expects a step in noise from $q^* = 1e$ outside the gap, to $q^* = 2e$ inside. This is because, inside the gap, single-electron processes are not allowed anymore, and only Andreev processes contribute to the noise. As shown in Fig. 4, our measurements are in qualitative agreement with this picture, both in NbSe$_2$ and FeTe$_{0.55}$Se$_{0.45}$. Outside the gap, our data follows the $q^* = 1e$ line, in agreement with Eq. (2). At the gap energy $\pm(\Delta_t + \Delta_s)$, a broadened step is visible in the extracted effective charge spectrum (Supplementary Fig. 5).

Interestingly, $q^*$ does not reach $2e$ inside the gap, but saturates at a value of $1.3e \sim 1.6e$ (Fig. 4b, d), despite a vanishing conductance within $\pm(\Delta_t + \Delta_s)$ in Figs. 2b and 3b. This is in contrast to the measurement at $B = 0$ T on FeTe$_{0.55}$Se$_{0.45}$, where the extracted $q^*$ reaches $1.97e$ at $\pm\Delta_t$ (see Supplementary Note 4). We hypnotize that the presence of the magnetic field leaves behind a small fraction of delocalized quasiparticles[49], which allows charges of $1e$ to tunnel. Even a very small fraction of quasiparticles will decrease the noise substantially, because for a given tunneling transparency $\tau$, the single-particle processes occur with probability $\tau$, while the Andreev processes occur with probability $\tau^2$ (see Supplementary Note 5 for an estimation of fractions). Future experiments at different magnetic fields, and mapping the exact spatial dependence of current noise around vortex cores are necessary to test this hypothesis relating to the effective charge away from the vortex core. In this study, we focus on the noise spectra in the centers of individual vortex cores.

To investigate the tunneling process into vortex bound states, we then measure the current noise at the vortex cores, first for NbSe$_2$. The experimental data in Fig. 4a show that noise in the core center follows the $1e$-noise behavior until reaching well within $\pm\Delta_t$. We observe a transition from $q^* = 1e$ to $q^* > 1e$ around $\pm 1.0$ meV, within which Andreev reflection at the tip side starts to dominate (Supplementary Fig. 5a). Nevertheless, the $q^*$ remains at $1.05e$ at $E = \pm\Delta_t$, where tunneling into the CdGM bound states occurs (Fig. 4b). As a comparison, we then measured the shot noise of tunneling into the vortex bound states in FeTe$_{0.55}$Se$_{0.45}$. To our surprise, the behaviors of noise and effective charge (Fig. 4c, d) at the vortex cores of FeTe$_{0.55}$Se$_{0.45}$ are very similar to those of NbSe$_2$, i.e., without any Andreev-reflection enhanced noise at $E = \pm\Delta_t$. We extract an effective charge $q^* = 0.99e$ into the zero-energy vortex bound states in FeTe$_{0.55}$Se$_{0.45}$, even closer to a single electron charge than that into CdGM bound states in NbSe$_2$.

**DISCUSSION**

We proceed to our discussion of which states are compatible with our results from noise measurements for FeTe$_{0.55}$Se$_{0.45}$. We start with YSR states, which have been observed to cause the zero-bias peak and the saturating conductance, and are present in FeTe$_{0.55}$Se$_{0.45}$[18]. YSR states originate from the resonant coupling between a superconductor and a magnetic impurity. One can tune the coupling by a local gate, such as a voltage-biased STM tip, and the energy levels of the YSR states shift correspondingly to the local field felt by the impurity[18]. Thus, we expect to see spatially dispersing in-gap conductance peaks when moving the tip away from the impurity. The spatial extent of YSR states on FeTe$_{0.55}$Se$_{0.45}$ is about 8 nm (Supplementary Fig. 6), comparable to the size of a vortex. One way is to examine the tunneling process into YSR states, which is expected to be dominated by Andreev reflection in the strong tunneling limit[17]. Naively, the difference compared to the tunneling process into CdGM states can be explained by the different natures of the two states: CdGM states live in the vortex core that extends throughout the superconductor, so that the tunneled electron can leave the superconductor via the one-dimensional vortex core; YSR states localize on the surface of a superconductor around a magnetic impurity, so the tunneled electron cannot go through the superconductor but Andreev-reflected as a hole. Therefore, we expect shot noise with an effective charge of $q^* = 2e$ when tunneling into YSR states[50].

To confirm this picture, we carry out tunneling conductance and noise measurements on the YSR states in FeTe$_{0.55}$Se$_{0.45}$ in the strong tunneling limit (see Supplementary Note 4). The YSR states appear as differential conductance peaks with a ring shape around an impurity site[18]. Our noise measurements (Supplementary Fig. 6) when tunneling into these YSR states show enhanced noise and $q^* \approx 2e$ and are indistinctive from those of tunneling into the bare superconductor. With their stark contrast to the noise and effective charge of $q^* = 1e$ measured for vortex bound states. This leads to the first conclusion of our paper: we can exclude YSR states as the origin of the zero-bias conductance peak.

We then turn to the possibility of Majorana bound states as the origin of the zero-bias peak, as put forward by Refs. [13,15,16]. Theoretical calculations[20–36] show that shot noise for tunneling into an isolated Majorana bound state vanishes for Majorana-induced Andreev reflection with unity probability, when the bias energy lies within the width of the Majorana bound state. For typical STM measurements where the bias energy is much larger than the intrinsic width of Majorana bound states[16], the tunneling shot noise is Poissonian, i.e., $q^* = 1e$. A second finding is thus that the shot-noise noise we measure is consistent with Majorana bound states.

However, we emphasize that our observation of the identical noise behavior and effective charge for CdGM bound states and the putative Majorana bound states clearly implies consistency with CdGM states as well, at least from a shot-noise point of view. The possible accompanying CdGM states[10] leading to the asymmetry in our differential conductance spectra, would not change the noise behavior but still lead to an effective charge $q^* = 1e$. No theoretical work has focused on tunneling processes into CdGM states. Thus, more theoretical and experimental studies are needed to understand the tunneling process into CdGM and Majorana bound states in vortices. Still, our work excludes YSR states, and therefore, taken together with high-resolution conductance measurements[15], points towards Majorana modes as likely candidate for the zero-bias peak.

More generally, our work represents a step towards determining the exact nature of a zero-energy state, following theoretical work for Majorana bound states in vortex cores and nanowires. We have measured local shot noise when tunneling into vortex bound states in individual vortices of NbSe$_2$ and FeTe$_{0.55}$Se$_{0.45}$. Using a superconducting tip, we demonstrate the feasibility of measuring shot noise even for states close to the Fermi level, which is usually overwhelmed by thermal noise. First, our data exclude YSR states as the origin of the zero-bias conductance peak at the vortex cores of FeTe$_{0.55}$Se$_{0.45}$. Second, while our data are in agreement with the theoretical prediction for Majorana bound states, we emphasize that we observe an identical shot noise behavior of topologically trivial CdGM bound states in NbSe$_2$.

More theoretical work, especially including a superconducting tip, might allow to gain more information from shot-noise studies. Such a model for the tunneling process from a superconducting tip into a Majorana bound state has already been developed[51], but only in the limit of low temperature. In the future, unambiguous identification of Majorana bound states by shot noise, might be possible in the low-bias limit where temperature and bias energy are both lower than the intrinsic width $\Gamma$ of the zero-bias state, i.e. $eV_{bias} \ll \Gamma$ and $k_BT \ll \Gamma$. This limit can be reached if shot noise measurements are enabled at milli-Kelvin temperatures. There, Majorana-induced resonant Andreev reflection leads to a vanishing shot noise because of unity transmission, which distinguishes itself from $q^* = 1e$ when only CdGM states exist in the vortex core. A further proposal in the low-bias limit[21] suggested a two-tip shot noise measurement setup on two different vortices: each tip tunnels into one localized vortex state, and a positive cross-correlated current noise is expected exclusively for Majorana states. One could further investigate the spin-resolved current-current correlation by a more sophisticated approach combining spin-polarized spectroscopy[53,54] with shot-noise measurements[55,56].

## METHODS

### Different Definitions of the Fano Factor

While the Fano factor was originally defined as the ratio between the variance and the mean value of a quantity, specific definitions vary in dealing with electrical current and its shot noise. An often applied definition of the Fano factor $F$ is the ratio between the shot noise power $S$ (precisely the Fourier transform of the current-current correlation function) and the Poisson noise $S_P$ due to independent single electrons[37],

$$F = S/S_P = S/2e|I|.$$

In some theory proposals[21,22,27,29] for shot noise of Majorana bound states, one different definition appears where the Fano factor is expressed as

$$F = P/e|I|,$$

where $P$ is the shot noise power (time-averaged current-current correlation function). Another definition of the Fano factor is expressed as the ratio between the differential noise power (the derivative of the time-averaged current-current correlation function with respect to the bias voltage) and the differential conductance[27,28],

$$F = \mathrm{d}P/\mathrm{d}V/(e \cdot \mathrm{d}I/\mathrm{d}V).$$

In the above definitions, however, the transmission of a single electron at a time is assumed. As a consequence, the correlation between them appears as sub- or super-Poissonian shot noise with $F < 1$ or $F > 1$, depending on the details of the transmission probabilities of the conducting channels. In this work, on the other hand, the charge transfer is the quantity of interest, and the STM junction is well in the single-channel, low-transmission regime (our highest tunnel conductance is 0.4 μS, yielding $\tau < 5.2 \times 10^{-3}$). In this regime, we include the possible correlation between charge carriers in the effective charge, $q^* = S/2|I|$, or, more precisely, following Eq. (2).

## Sample preparation and STM measurements

The FeTe$_{0.55}$Se$_{0.45}$ single crystals with a transition temperature $T_C = 14.5$ K were grown using the Bridgman method. The 2H-NbSe$_2$ samples ($T_C = 7.2$ K) are purchased from HQ Graphene. The samples with a thickness of ~ 0.5 mm are cleaved in an ultrahigh vacuum at ~ 30 K and immediately inserted into a customized STM (USM-1500, Unisoku Co., Ltd). All measurements are performed in a cryogenic vacuum at a base temperature of $T = 2.3$ K. We perform scanning tunneling spectroscopy using standard lock-in techniques without the feedback loop enabled. A bias voltage modulation at a frequency of 887 Hz with an amplitude of 100 μV (for maps around vortex) or 50 μV (for high-resolution point spectra) is applied. The resulting differential conductance (d$I$/d$V$) values are normalized by setup conductance $I_{set}/V_{set}$. Prior to all the measurements, a Pt-Ir tip is made superconducting by indenting it into a clean Pb(111) surface. Our superconducting tip exhibits a critical field of about 0.7 T, deducted from differential conductance measurements in different magnetic fields on an atomically flat Au(111) surface (see Supplementary Note 1 for details).

## Noise measurements

We perform noise spectroscopy at a constant junction resistance $R_J$ in a slow feedback loop (see Supplementary Note 6 for details) when varying the bias voltage $V_{bias}$ (and hence tunneling current $I = V_{bias}/R_J$) using our custom-built cryogenic megahertz amplifier developed recently. Because the Josephson tunnel junction with a low $R_J$ may couple to its environment[41,52], which affects the measured noise, we keep our $R_J > 2.5$ MOhm and $V_{bias} > 0.2$ mV, where the Andreev-reflection enhanced conductance at $\pm\Delta_t$ and the environmental coupling effect is negligible. The amplifier consists of an LC tank circuit and a high-electron-mobility transistor that converts the current fluctuations in the junction into voltage fluctuations across a 50 Ohm line, as described in detail elsewhere[40]. To extract the effective charge transferred in the junction we follow a similar procedure as described in Refs. [41,42].

The measured total voltage noise is

$$S_V^{\mathrm{meas}}(\omega, V) = G^2 |Z_{\mathrm{tot}}|^2 S_I,$$

where $G$ is the total gain calibrated by noise spectrum at a high bias (see Supplementary Fig. 9), and $S_I$ is the total current noise

$$S_I = 2q^* I \coth\left(\frac{q^* V}{2k_B T}\right) + \frac{4k_B T R_{\mathrm{res}}}{|Z_{\mathrm{tot}}|^2} + S_{\mathrm{amp}}.$$

The first term is the junction noise from Eq. (2), the second is the thermal noise originating from the resistive part $R_{\mathrm{res}}$ of the LC tank circuit, and $S_{\mathrm{amp}}$ is the intrinsic current noise of our amplifier.

As the first step of the procedure, we measure the background noise by retracting the tip out of tunneling ($I=0$ so the first term vanishes), which gives $4k_B T R_{\mathrm{res}}/|Z_{\mathrm{res}}|^2 + S_{\mathrm{amp}}$, where $Z_{\mathrm{tot}} = Z_{\mathrm{res}}$ because the junction is an open circuit. Then we measure noise in tunneling and subtract it by the background noise (for low $R_J$ we also consider $Z_{\mathrm{tot}}$ in the second term as $Z_{\mathrm{res}}$ in parallel with $R_J$). Thus, the current noise data plotted in Fig. 4a, c consists only of the noise from the junction. Finally, we extract the $q^*$ value at each bias by numerically solving Eq. (2).

## DATA AVAILABILITY

## CODE AVAILABILITY

The code used for this project is available upon request to the authors.

## ACKNOWLEDGMENTS

We acknowledge C. W. J. Beenakker, C. J. Bolech, T. Hanaguri, T. Machida, D. K. Morr, F. von Oppen, and J. van Ruitenbeek for valuable discussions.

**Funding:** This work was supported by the European Research Council (ERC StG SpinMelt). K.M.B. was supported by the Netherlands Organization for Scientific Research (NWO Veni grant VI.Veni.212.019). D.Cho was supported by the National Research Foundation of Korea (NRF) funded by the Korea government (MSIT) (grants 2020R1C1C1007895 and 2017R1A5A1014862) and the Yonsei University Research Fund (grant 2019-22-0209). G.G. was supported by the Office of Basic Energy Sciences, Materials Sciences and Engineering Division, US Department of Energy (DOE) under contract number de-sc0012704.

## AUTHOR CONTRIBUTIONS

J-F.G., K.M.B., D.Cha., D.Cho, W.O.T., T.B., and J.N. performed the experiments and analyzed the data. G.G. grew and characterized the $FeTe_{0.55}Se_{0.45}$ samples. All authors contributed to the interpretation of the data and writing of the manuscript. M.P.A. supervised the project.

## COMPETING INTERESTS

Authors declare that they have no competing interests.

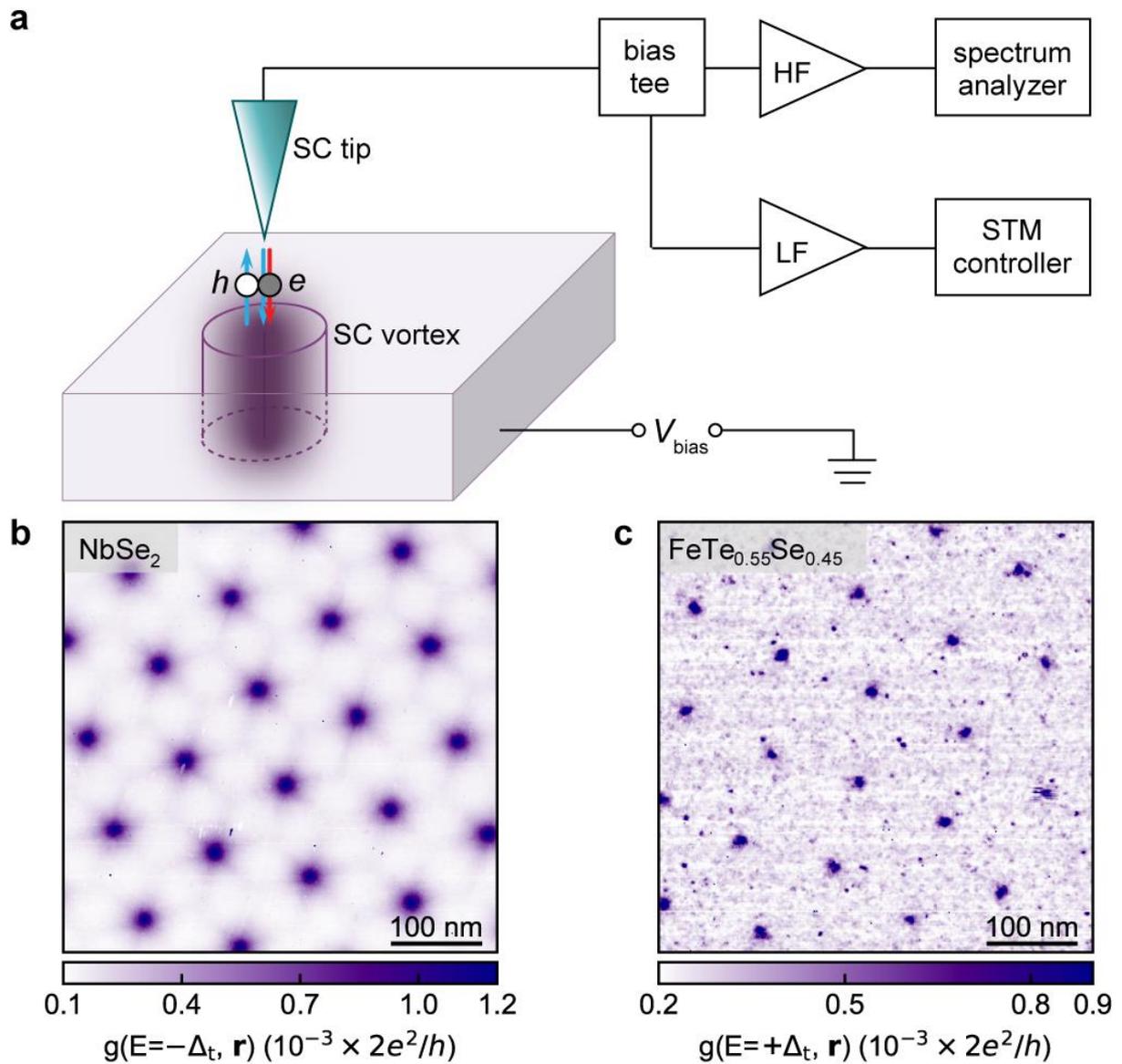

**Fig. 1 Local tunneling shot noise measurements of vortex bound states. a** Schematic illustration of the scanning tunneling noise microscope setup. A bias voltage ($V_{bias}$) is applied between the superconducting (SC) tip and sample, while the cylinder represents a SC vortex. If the tunneling process is a single-electron (gray) into vortex bound states (red arrow), an effective charge $q^* = 1e$ is transferred from the tip to the vortex. When Andreev reflection takes place (blue arrows), a hole (white) is reflected, and the effective charge doubles $q^* = 2e$. HF and LF stand for the high- and low-frequency amplifier, respectively. STM, scanning tunneling microscope. **b,c,** Full flux quantum ($h/2e$) vortex lattice in NbSe$_2$ (**b**) and FeSe$_{0.55}$Te$_{0.45}$ (**c**) revealed by spatially resolved differential conductance at a magnetic field of 0.1 T. Setup conditions: b, $V_{set}$ = -5 mV, $I_{set}$ = 200 pA; c, $V_{set}$ = 10 mV, $I_{set}$ = 250 pA.

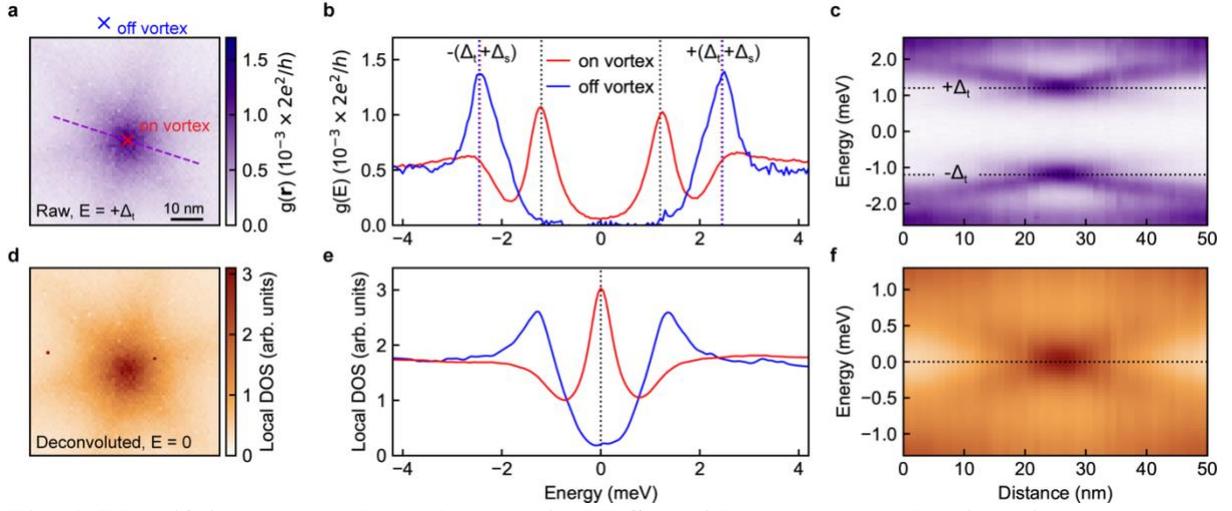

**Fig. 2 Identifying vortex bound states in NbSe₂ with a superconducting tip. a** Spatially resolved differential conductance around an individual vortex at the energy $E = +\Delta_t$, where $\Delta_t$ and $\Delta_s$ stand for the superconducting gaps of the tip and sample, respectively. **b** High-resolution differential conductance spectra acquired at the two locations marked by the crosses in **a**: the center of the vortex core (red) and off vortex (blue). The off-vortex location is 60 nm away from the core center, in the midpoint between two neighboring vortices of Fig. 1b. The red and blue dashed lines indicate the coherence peaks. **c** Differential conductance spectra along the dashed line (purple) in **a** showing the spatial dispersion of the vortex bound states. The gray dashed lines in **b** and **c** indicate the peaks at $\pm\Delta_t$ where tunneling into the vortex bound states occurs at the core center. **d**-**f**, Local density of states (DOS) plots corresponding to **a**-**c**, after deconvolution using the tip DOS. The vortex bound states are indicated by the peak in the local DOS at around zero energy (gray dashed line). **b**, **e** and **c**, **f** share the same horizontal axes. Setup conditions: $V_{set} = 5$ mV, $I_{set} = 200$ pA.

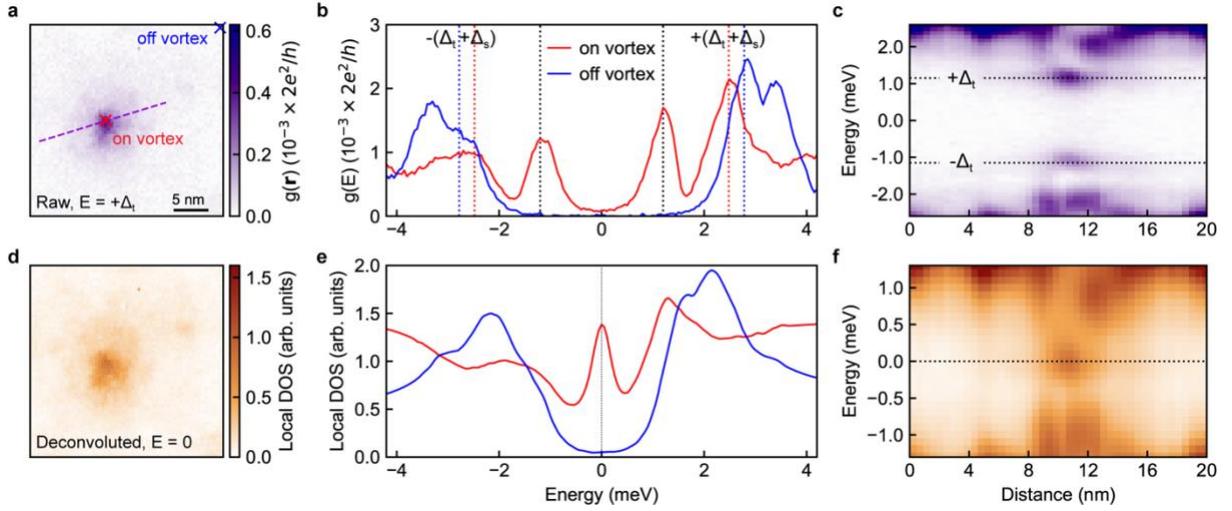

**Fig. 3 Identifying the putative Majorana bound state in FeTe$_{0.55}$Se$_{0.45}$. a** Spatially resolved differential conductance around an individual vortex at the energy $E = +\Delta_t$, where $\Delta_t$ and $\Delta_s$ stand for the superconducting gaps of the tip and sample, respectively. **b** High-resolution differential conductance spectra acquired at the two locations marked by the crosses in **a**: the center of the vortex core (red) and off vortex (blue). The red and blue dashed lines indicate the coherence peaks. **c** Differential conductance spectra along the dashed line (purple) in **a** showing the spatial extent of the zero-energy bound state. The gray dashed lines in **b** and **c** indicate the peaks at $\pm\Delta_t$, where tunneling into the putative Majorana bound state occurs. **d-f** Local density of states (DOS) plots corresponding to **a-c**, after deconvolution using the tip DOS. The putative Majorana bound state is indicated by the peak in the local DOS at zero energy (gray dashed line). **b**, **e** and **c**, **f** share the same horizontal axes. Setup conditions: **a,c**, $V_{set} = 10$ mV, $I_{set} = 250$ pA; **b**, $V_{set} = 5$ mV, $I_{set} = 250$ pA.

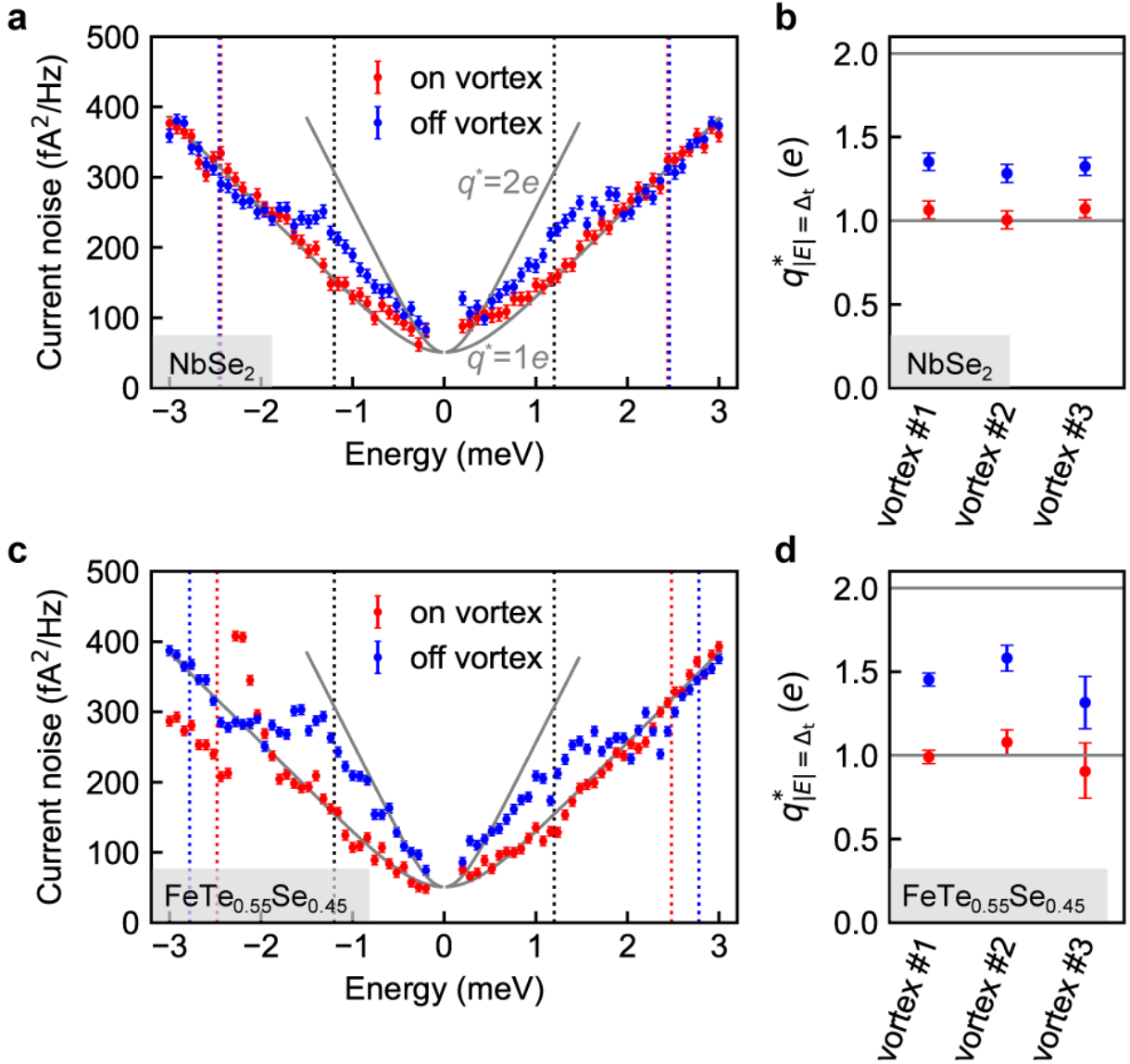

**Fig. 4 Local noise spectroscopy on and off vortices in NbSe₂ and FeTe₀.₅₅Se₀.₄₅. a,c** Current noise spectra in the tunnel junction (with a resistance $R_J$ = 2.5 MOhm) taken on (red) and off (blue) the vortex shown in Fig. 2a for NbSe₂ and Fig. 3a for FeTe₀.₅₅Se₀.₄₅, respectively. The locations of these spectra are marked by the crosses in Figs. 2a and 3a with the same colors. Gray curves are the expected noise from Eq. 2 with an effective charge $q^*$ of 1$e$ and 2$e$ at $T$ = 2.3 K. The dashed lines in a and c are replicated from Fig. 2b and Fig. 3b, respectively, serving as guides for the coherence peaks (red and blue) and the bound states (gray). The error bars are determined by the fluctuation of the current noise in time before each experiment, yielding a standard deviation of 9.25 and 6.77 fA²/Hz for **a** and **c**, respectively. **b,d** Effective charge $q^*$ derived by numerically solving Eq. 2 at the energy $E = \pm\Delta_t$ on (red) and off (blue) vortex for three different vortices in NbSe₂ and FeTe₀.₅₅Se₀.₄₅, respectively. The error bars are determined by the standard deviation of the extracted $q^*$ (Supplementary Figs. 3-5) in the energy ranges ($\Delta_t \pm 0.1$ meV) and -($\Delta_t \pm 0.1$ meV).

# Single-electron charge transfer into putative Majorana and trivial modes in individual vortices

*Jian-Feng Ge,[1] Koen M. Bastiaans,[1, 2] Damianos Chatzopoulos,[1] Doohee Cho,[3] Willem O. Tromp,[1] Tjerk Benschop,[1] Jiasen Niu,[1] Genda Gu,[4] Milan P. Allan[1]\**

[1] *Leiden Institute of Physics, Leiden University, 2333 CA Leiden, The Netherlands*
[2] *Department of Quantum Nanoscience, Kavli Institute of Nanoscience, Delft University of Technology, 2628 CJ Delft, The Netherlands*
[3] *Department of Physics, Yonsei University, Seoul 03722, Republic of Korea*
[4] *Condensed Matter Physics and Materials Science Department, Brookhaven National Laboratory, Upton, NY, 11973, USA*

*Corresponding author. Email: allan@physics.leidenuniv.nl

## Supplementary Note 1: Characterization of the superconducting tip

We decorate a mechanically ground Pt-Ir tip with Pb microcrystals by indenting into a Pb(111) single crystal, cleaned beforehand by standard Ar$^+$ sputtering cycles, until a superconductor-insulator-superconductor (SIS) tunnel junction is established (Supplementary Fig. 1a). The energy resolution is estimated to be 0.25 meV from the full width at half maxima of the sharp coherence peaks in Supplementary Fig. 1a.

Then we perform tunneling spectroscopy with this tip on a clean Au(111) surface at various magnetic fields, as shown in Supplementary Fig. 1b. The differential conductance measured by tunneling spectroscopy is expressed by

$$\frac{\mathrm{d}I(V)}{\mathrm{d}V} \propto \int \mathrm{d}E\ N_s(E) \frac{\partial}{\partial V} \{N_\mathrm{t}(E+eV)[f(E,T) - f(E+eV,T)]\}, \qquad (1)$$

where $E$ is energy, $N_s$ ($N_\mathrm{t}$) is the density of states (DOS) in the sample (tip), and $f(E, T) = [1+\exp(E/k_\mathrm{B}T)]^{-1}$ is the Fermi function ($k_\mathrm{B}$ being the Boltzmann constant). We model the tip DOS by the Dynes function[1]

$$N_\mathrm{t}(E, V, \Gamma) = \mathrm{Re}\left[\frac{E+i\Gamma}{(E+i\Gamma)^2 - \Delta_\mathrm{t}^2}\right], \qquad (2)$$

where $\Delta_\mathrm{t}$ is the superconducting energy gap of the tip and $\Gamma$ is the phenomenological broadening parameter (without thermal broadening). The values for $\Delta_\mathrm{t}$ and $\Gamma$ are extracted by fitting each spectrum to Supplementary Equation 1, assuming a constant $N_s$ for Au(111) in the energy range from -10 meV to +10 meV. The fit results are summarized in Supplementary Fig. 1c, showing a critical field of 0.7 T, about 7 times larger than that of the bulk Pb.[2]

## Supplementary Note 2: Deconvolution of the conductance spectra taken with a superconducting tip

The consequence of using a superconducting tip with an energy gap $\Delta_\mathrm{t}$, as illustrated by Supplementary Fig. 2a, is the resonance tunneling when either of the gap edges of the tip DOS aligns with the zero-energy state (ZES). Thus, the bound state appears as peaks at $\pm\Delta_t$ in the measured differential conductance spectra. Deconvolution of a differential conductance spectrum is necessary to recover the appearance of the bound state at zero energy in the sample DOS, the same as what one would expect for a spectrum taken with a normal-metal tip.[3, 4] We follow the deconvolution algorithm described in Refs. [5, 6] to extract the sample DOS $N_s$ in Supplementary Equation 1. We use the fit results at $B = 0.1$ T in Supplementary Fig. 1c for the tip DOS $N_t$ in the deconvolution.

To determine the energy of the vortex bound states in FeTe$_{0.55}$Se$_{0.45}$, in Supplementary Fig. 2b, c we stack the raw and deconvoluted spectra for the line cut in Fig. 3c, f, respectively. We fit each peak (at $\pm\Delta_t$ for Supplementary Fig. 2b and at 0 meV for Supplementary Fig. 2c) with a Lorentzian function, and plot the peak energy as a function of the position along the line in Supplementary Fig. 2d. We note an additional broadening of the coherence peaks and the zero-bias peak in the deconvoluted DOS (Fig. 3e) compared to the peaks in the raw spectrum in Fig. 3b, due to the need in the deconvolution algorithm to remove oscillatory errors (we chose the optimized value for the control parameter $\gamma = 5.0$ defined in Ref. [5]). Nevertheless, the peak center locates at $0 \pm 50$ μeV, confirming them as zero-energy states. On the other hand, from the amplitudes of the Lorentzian fit (Supplementary Fig. 2e), we find the zero-energy state has roughly an exponential decay in DOS, with a decay length of ~ 4.0 nm on both sides of the core center.

**Supplementary Note 3: Differential conductance and noise spectroscopy on different vortices**

We present the full dataset of both differential conductance and noise spectroscopy performed on all vortices in Supplementary Figs. 3 (for NbSe$_2$) and 4 (for FeTe$_{0.55}$Se$_{0.45}$), except the ones shown in Figs. 2 to 4. For each material, all vortices exhibit similar behavior as illustrated in the main text, especially a zero-bias peak in the DOS after deconvolution and 1$e$-noise on vortex. We show in Supplementary Figs.5a and b the extracted effective charge $q^*$ as a function of bias energy for the noise spectra in Figs. 4a and c, respectively. The error bars of $q^*$ in Fig. 4b (4d) are extracted from Supplementary Figs.5a, 3h, 3p (Supplementary Figs. 5b, 4h, 4p), within an energy window of 0.2 meV centered at $\pm\Delta_t$. A bigger error in $q^*$ for vortex #2 (and #3) in Fig. 4d is due to a higher junction resistance of 5 MOhm (and 10 MOhm) used for the noise spectroscopy measurements, leading to reduced absolute values of tunnel current and its noise.

**Supplementary Note 4: Scanning noise spectroscopy at zero field around a YSR impurity**

The results of our measurements on FeTe$_{0.55}$Se$_{0.45}$ at zero magnetic field are shown in Supplementary Fig. 6. Here we observe the Yu-Shiba-Rusinov (YSR) bound states as a ring in the differential conductance map (Supplementary Fig. 6b). The YSR states lead to a negative differential conductance in the spectrum (Supplementary Fig. 6d) because of the convolution of a sharp in-gap resonance peak and superconducting tip DOS as we observed in a previous study.[6] The noise spectrum measured *away* from the impurity site (Supplementary Fig. 6e) shows clear transitions from $q^* = 1e$ line to $q^* = 2e$ line with onsets at $\pm(\Delta_t + \Delta_s)$, indicating a dominating Andreev reflection inside the gap. This noise behavior is similar to what we observed before on Pb(111) surface with a superconducting tip.[7] We extract the effective charge $q^*$ in Supplementary Fig. 6f by numerically solving Eq. 2 in the main text. We observe a narrower step of $q^*$ from 1$e$ outside the gap to 1.97$e$ at $E = \pm\Delta_t$, compared to the broader transition from $1e$ to a plateau of $1.3e\sim1.6e$ in Supplementary Fig. 5. The value of $q^*$ so close to $2e$ in Supplementary Fig. 6f indicates the tunneling current originates purely from Andreev reflection in the SIS junction, whereas $q^*$ short of $2e$ in Supplementary Fig. 5 off vortex implies that the tunneling process is not purely Andreev reflection, and that a contribution from $1e$-charge tunneling coexists.

A previous study[8] demonstrated, by both theory and experiment, that Andreev process dominates single-particle tunneling into YSR states in the strong tunneling limit $\Gamma_1 \ll \Gamma_t$, where $\Gamma_t$ is the tunneling rate and $\Gamma_1$ is the threshold rate for quasiparticles to be excited into the continuum. For our experiment on FeTe$_{0.55}$Se$_{0.45}$, we estimate $\Gamma_1 = 1.1$ μeV from Eq. (S49) of Ref. [8], using $\Delta_s = 1.5$ meV, $T = 2.3$ K, and the YSR energy of 0.3 meV. This yields a threshold current of ~ 90 pA from Eq. (S65) of Ref. [8]. We expect that, in this limit, the effective charge when tunneling into YSR states is identical to that into the bare superconductor at a bias energy inside the gap.

We have verified in the strong tunneling limit (with a current of 800 pA) that the YSR state does not cause a spatial difference in shot noise, as shown by the noise map in Supplementary Fig. 6c, compared to the prominent ring feature in the differential conductance map taken in the same field of view and at the same bias voltage (Supplementary Fig. 6b). Therefore, we exclude YSR states as the origin of the zero-energy bound states in the vortex cores of FeTe$_{0.55}$Se$_{0.45}$.

Note that there is no contradiction between Ref. [9] and our data for YSR states. For tunneling into the bare superconductor NbSe$_2$, Ref. [9] measures $1e$ noise outside the gap but lacks of noise data inside the gap because of the low current (~20 pA), for which they could not detect the corresponding shot noise accurately. In principle, considering a normal tip used in this case, inside the gap $q^*$ increases gradually to $2e$ [10]. For tunneling into YSR states, Ref. [9] elucidates that Andreev reflection requires particle-hole symmetric resonances. The smaller of the asymmetric resonances leads to a deficiency of providing particle (or hole) for the Andreev ($2e$) process, so that tunneling into the excess hole (or particle) component in the larger resonance has to be mediated by an inelastic quasiparticle relaxation process. In our experiment this inelastic quasiparticle relaxation ($1e$) process is strongly suppressed because the YSR states in FeTe$_{0.55}$Se$_{0.45}$ are almost symmetric in amplitude and our noise measurements were carried out in the strong tunneling regime. In summary, both our data and Ref. [9] reach the conclusion that tunneling into YSR states, when inelastic quasiparticle relaxation can be neglected, gives $2e$ noise.

## Supplementary Note 5: The effective charge when single-particle and Andreev processes both contribute

In this section, we calculate, based on an empirical model, the effective charge when both Andreev reflection and quasiparticle of $1e$ tunneling contribute to the total current. Away from the vortex, the tunnel junction is similar to an SIS junction, as shown by comparing the spectra in Fig. 2b and Fig. 3b and the zero-field spectrum in Supplementary Fig. 6d. Especially at the bias energy $E = \pm\Delta_t$, the deconvolution yields a vanishing density of states of the sample (Figs. 2e and 3e). Therefore, the tunneling process for this tunnel junction is expected to be dominated by Andreev reflection that transfers a charge of $2e$ per event. We then introduce a fraction of the $1e$-charge tunneling process,[11] which contributes to current and noise but has no correlation with those of the Andreev process. For a given tunneling transparency $\tau \ll 1$, the current contributions for the single-particle processes ($I_{1e}$) and the Andreev processes ($I_{2e}$) are proportional to $\tau$ and $\tau^2$, respectively,[12]

$$I_{ne} \propto n\tau^n/4^{n-1}, \, n = 1, \, 2. \tag{S3}$$

The prefactors are related to the integrated density of states, and here we assume an empirical prefactor $y$ for quasiparticle contribution $I_{1e}$ and $1-y$ for $I_{2e}$ to have a conserved total integrated density of states. The total current is $I = I_{1e} + I_{2e}$. As $I_{1e}$ and $I_{2e}$ are assumed to be independent, the total current noise is the sum of both contributions,

$$S = 2eI_{1e}\coth(eV/2k_BT) + 2\cdot2e\cdot I_{2e}\coth(2eV/2k_BT), \tag{S4}$$

where the double-charge ($2e$) transfer is taken into account in the Andreev contribution (the second term). Then we extract numerically the (total) effective charge $q^*$ by Eq. 2.

Supplementary Figure 7 plots $q^*$ as a function of the fraction of quasiparticle contribution for different junction resistance we used in noise measurements. When $I_{1e}/I = 0$, i.e., no single-particle process contributes, $q^* = 2e$ as expected from purely Andreev reflection. Conversely when $I_{1e}/I = 100\%$, only single-particle process contributes, yielding $q^* = 1e$. For values of $I_{1e}/I$ in between 0 and 100%, we find a quick reduction of $q^*$ even when a very small fraction of quasiparticle contribution exists (note the logarithmic scale of the horizontal axis). For example, for $R_J = 2.5$ MOhm, 0.02% of quasiparticle contribution reduces $q^*$ to $1.92e$, while 3.3% of quasiparticle contribution reduces already reduces $q^*$ to $1.07e$.

## Supplementary Note 6: Transparency of the tunnel junction during noise spectroscopy

In differential conductance (Figs. 2 and 3) and noise (Fig. 4) spectroscopy, we use different setup conditions in terms of feedback control of the tip. Specifically, in differential conductance measurements, as the protocols are conventionally applied, feedback is disabled during voltage sweeps (spectroscopy). However, in noise spectroscopy, in order to have optimal junction stability, we enable a slow feedback to maintain a constant junction resistance (i.e., changing the bias voltage and current setpoint for each point in a sweep), except at the zero-bias point where feedback has to be disabled. As already presented in Fig. 2b, the differential conductance of an SIS junction varies by an order of magnitude during a sweep, the transparency of the junction, if feedback is enabled, could also vary considerably. The transparency $\tau$, which is assumed to be in the $\tau \ll 1$ limit for current noise expressions in the main text, has a significant influence on the resulted noise when it becomes comparable to unity.[9] In Supplementary Fig. 8a we compare the differential conductance taken with feedback disabled and enabled, at different junction resistance (thus the setup $\tau$). While outside the gap the conductance measured both ways is almost identical, a drastic difference in conductance develops in the gap because of a vanishing quasiparticle density of states. The ratio of the feedback-on conductance over the feedback-off conductance $g_{on}/g_{off}$ indicates the enhancement of transparency from $\tau = (R_J G_0)^{-1}$ from an Ohmic current-voltage relation, where $G_0 = 2e^2/h = 77.5$ μS is the conductance quantum ($h$ being the Planck constant). In fact, the conductance ratio $g_{on}/g_{off}$ is barely dependent on $R_J$, as shown in Supplementary Fig. 8b. To estimate the highest $\tau$ inside the gap throughout our noise measurements around vortices, we approximate $\tau = (R_J G_0)^{-1}$ for the setup bias, where $g_{on}/g_{off}$ is close to 1. For simplicity, we model the sample DOS $N_s$ also by the Dynes function (Supplementary Equation 2) and calculate the feedback-off conductance $g_{off}$ by Supplementary Equation 1. In this case, the conductance ratio $g_{on}/g_{off}$ is equal to the current ratio $I_{on}/I_{off}$. Therefore, we first integrate $g_{off}$ to get $I_{off}$, and then we obtain the ratio $g_{on}/g_{off} = I_{on}/I_{off} = VR_J/I_{off}$. Using this simple model, we can simulate the ratio $g_{on}/g_{off}$ (the green line in Supplementary Fig. 8b), in good agreement with the experimental results. From Supplementary Fig. 8 we have confirmed that in our measurement conditions, enabling feedback has a marginal effect on the noise, as the transparency $\tau$ stays below 0.16, which is still in the $\tau \ll 1$ limit.

## Supplementary Note 7: Different dispersions of CdGM states in NbSe$_2$ and FeTe$_{0.55}$Se$_{0.45}$

For a conventional s-wave BCS superconductor, CdGM states have been extensively studied theoretically (e.g., Ref. S11): the CdGM states have an increasing angular momentum when moving a distance $r$ away from the vortex core. As a consequence, the majority of the CdGM states that contributes to the differential conductance have an energy $E_p$ approximately proportional to $k_F \cdot r$, where $k_F$ is the Fermi wavevector. In addition, at $E_p$ the differential conductance maximum decays exponentially in $r$ on a length scale of coherence length $\xi$. Therefore, the dispersion profile of CdGM states depends crucially on two material parameters $k_F$ and $\xi$ (see Supplementary Table 1 for their values of NbSe$_2$ and FeTe$_{1-x}$Se$_x$). For NbSe$_2$, both parameters are larger, and the dispersion is measurable by STM. However, for FeTe$_{1-x}$Se$_x$, $k_F$ is one order of magnitude smaller, so $E_p$ changes much more slowly with $r$; meanwhile $\xi$ is also smaller, resulting in a vanishing amplitude in differential conductance before $E_p$ changes significantly. Therefore, observation of dispersing states indicates a CdGM origin, but non-dispersing states cannot exclude a CdGM origin.

## Supplementary Table 1. Fermi wavevector and coherence length of NbSe$_2$ and FeTe$_{1-x}$Se$_x$.

| Material | NbSe$_2$ | FeTe$_{1-x}$Se$_x$ |
|---|---|---|
| Fermi wavevector $k_F$ | 0.5~1 Å$^{-1}$ [12] | 0.07~0.12 Å$^{-1}$ [13] |
| Coherence length $\xi$ | 12 nm [14] | 3 nm [15] |

## Supplementary References

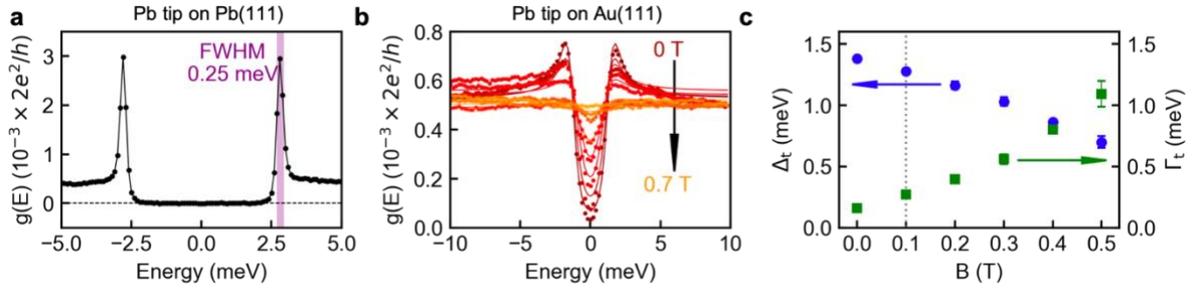

**Supplementary Figure 1. Characterization of the superconducting tip. a** Differential conductance spectrum after indenting the superconducting tip in a Pb(111) surface. FWHM, full-width at half-maximum. **b** Magnetic-field dependence of the differential conductance spectra (dots) of the tip in (A) on an Au (111) surface. Solid lines show the fit by Supplementary Eq. 1 to each spectrum. The magnetic field is increased from 0 T (dark red) to 0.7 T (yellow) with a 0.1 T interval. **c** Fit parameters $\Delta_t$ (left axis) and $\Gamma$ (right axis) as a function of the magnetic field. Error bars stand for uncertainties extracted from the fit. The spectra at $B = 0.6$ T and $B = 0.7$ T are almost flat, yielding large error bars of $\Delta_t$ and $\Gamma$ in the fit results. Setup conditions: **a**, $V_{set} = 5$ mV, $I_{set} = 200$ pA; **b**, $V_{set} = 10$ mV, $I_{set} = 400$ pA.

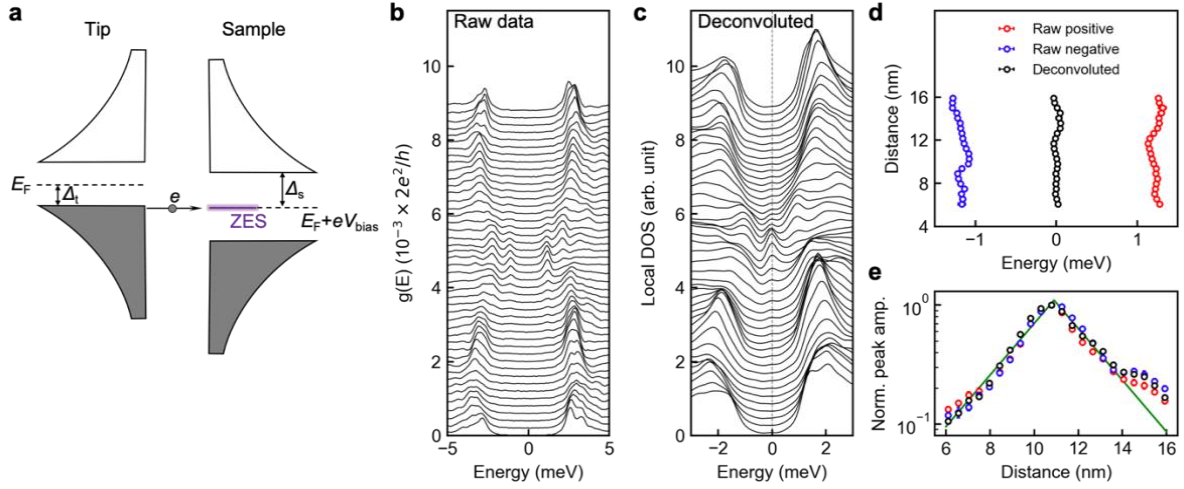

**Supplementary Figure 2. Details of the structure of the zero-energy vortex bound states in FeTe$_{0.55}$Se$_{0.45}$.** **a** An energy diagram showing the tunneling process from a superconducting tip to the zero-energy state (ZES) in FeTe$_{0.55}$Se$_{0.45}$ with a sample bias $eV_{bias} = -\Delta_t$. Both the tip and sample are superconducting with a gap size of $\Delta_t$ and $\Delta_s$, respectively. The gray (white) area denotes the occupied (empty) states with a diverging DOS near gap edges. At this bias the diverging quasiparticle DOS leads to a maximal probability tunneling into the ZES (purple), resulting in enhanced differential conductance. $E_F$, Femi level. **b,c** Differential conductance (**b**) and corresponding deconvoluted local DOS (**c**) spectra for the line cut images shown in Figs. 3c and 3f, respectively. Spectra are shifted with a spacing of 0.2 for clarity. A Lorentzian fit is carried out to each peak inside the gap of each spectrum in **b** and **c**. **d,e** The peak energy (**d**) and (**e**) normalized peak amplitude (Norm. peak amp.) for points at different distances in the line cuts. The peak amplitude is normalized by the maximum of each series: red (blue) for the peak with a positive (negative) energy in **b**, and black for the peak near zero energy in **c**. The greed lines show a symmetric exponential decay around 10.9 nm (location of the vortex core center) with a decay length of 4.0 nm. Note the logarithmic scale of the vertical axis in **e**.

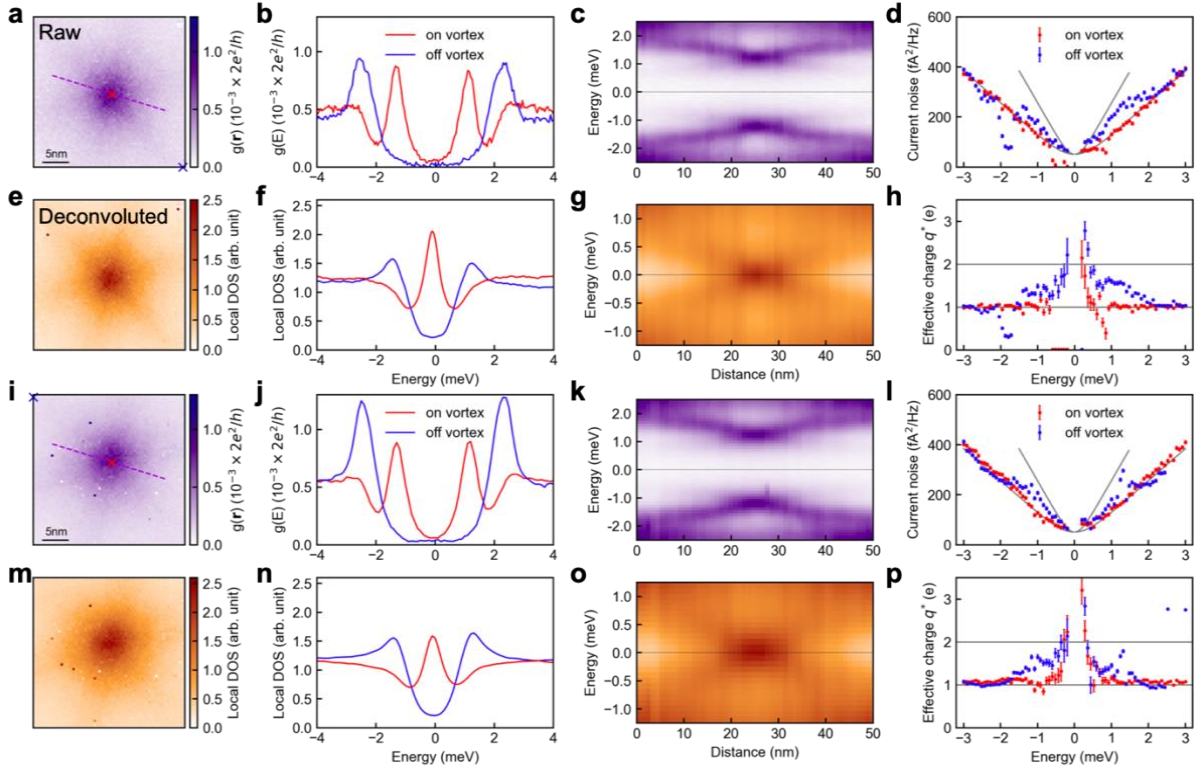

**Supplementary Figure 3. Differential conductance and noise spectroscopy on different vortices in NbSe₂. a-g** Differential conductance (**a-c** and **e-g**) and shot noise (**d**) measurements same as shown in Fig. 2 and Fig. 4a, respectively, for vortex #2 in Fig. 4b. **h,** The effective charge as a function of bias energy extracted from **d**. **i-p** Same as **a-h** for vortex #3 in Fig. 4b. Error bars in **d** and **l** are determined by the fluctuation of the current noise in time, yielding a standard deviation of 9.25 fA²/Hz. Error bars in **h** and **p** correspond to numerical solutions of Eq. 2 in the main text, using values of upper and lower bounds indicated the error bars in **d** and **l**, respectively.

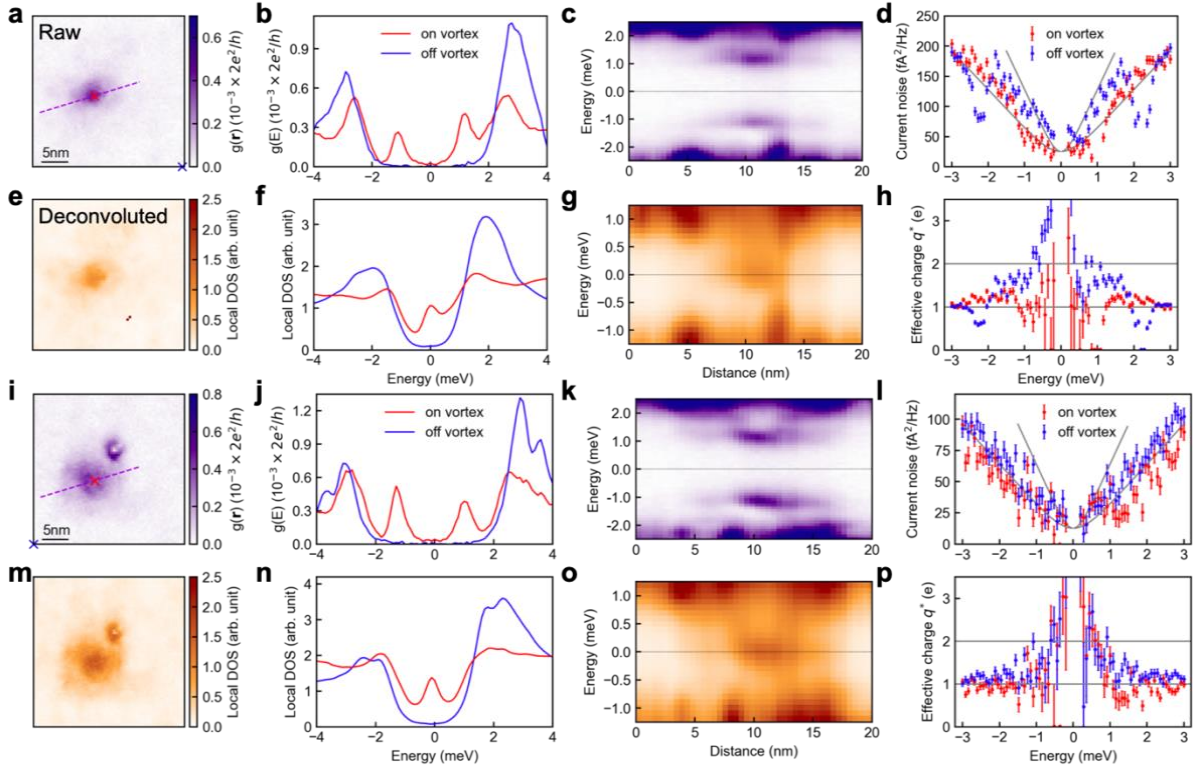

**Supplementary Figure 4. Differential conductance and noise spectroscopy on different vortices in FeTe$_{0.55}$Se$_{0.45}$. a-g** Differential conductance (**a-c** and **e-g**) and shot noise (**d**) measurements same as shown in Fig. 3 and Fig. 4c, respectively, for vortex #2 in Fig. 4d, except for a junction resistance $R_J$ = 5 MOhm used in noise spectroscopy. **h**, Effective charge as a function of bias energy extracted from **d**. **i-p** Same as **a-h** for vortex #3 in Fig. 4d, except for a junction resistance $R_J$ = 10 MOhm used in noise spectroscopy. Error bars in **d** and **l** are determined by the fluctuation of the current noise in time, yielding a standard deviation of 6.77 fA$^2$/Hz. Error bars in **h** and **p** correspond to numerical solutions of Eq. 2 in the main text, using values of upper and lower bounds indicated the error bars in **d** and **l**, respectively. Note that a YSR impurity is observed, as indicated by the small ring feature next to the vortex in **i**.

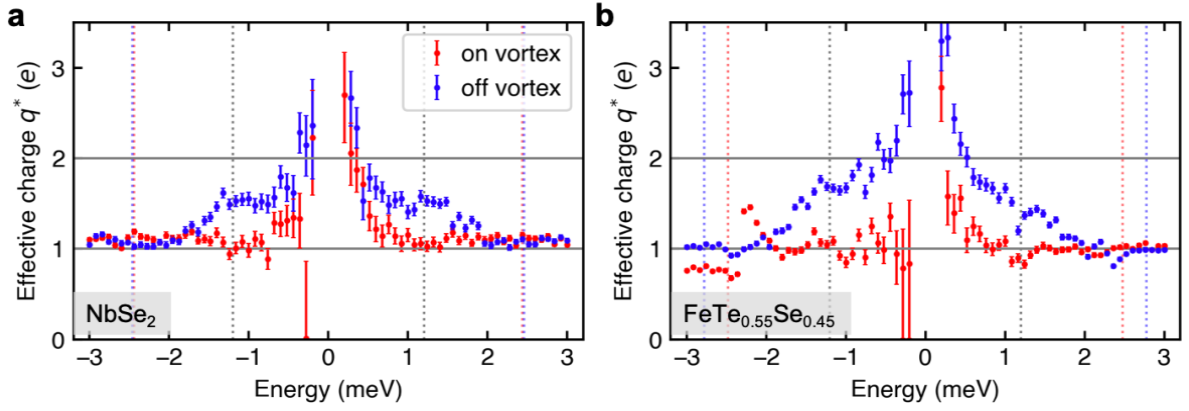

**Supplementary Figure 5. Effective charge spectra on and off vortex in NbSe₂ and FeTe₀.₅₅Se₀.₄₅. a,b** The effective charge numerically extracted from Fig. 4a and Fig. 4c, respectively, by Eq. 2 in the main text. Error bars correspond to numerical solutions using values of upper and lower bounds indicated the error bars in Fig. 4a and Fig. 4c, respectively. A further increase of $q^*$ above within ±0.4 meV is caused by an increasing Andreev (and possible multiple Andreev) contribution because of a vanishing tip DOS. Unfortunately, the uncertainty also increases quickly for the numerical solution within this range due to the divergence of the coth function when $V_{bias} \rightarrow 0$ in Eq. 2 in the main text.

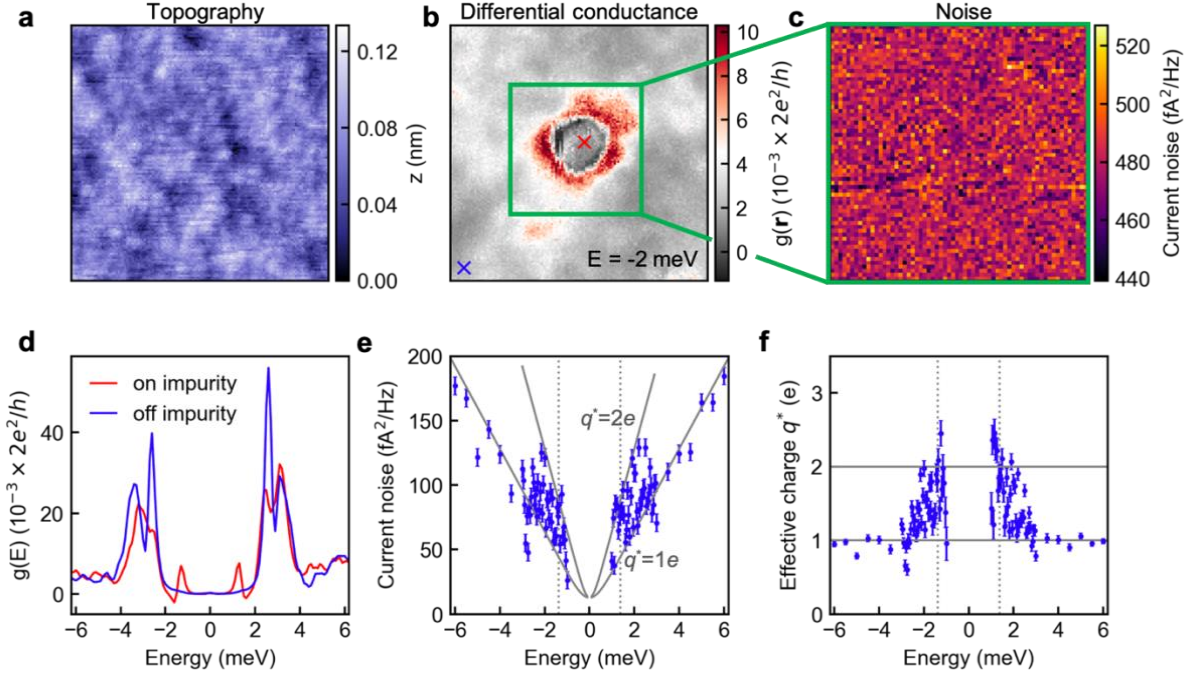

**Supplementary Figure 6. Noise measurements near a YSR impurity at zero magnetic field. a,b** STM topography (**a**) and differential conductance (**b**) for bias $V_{bias}$ = -2 mV for the same field of view (20 nm × 20 nm). A ring feature with enhanced (reduced) conductance outside (inside) indicates that resonant tunneling occurs when approaching the impurity. **c** Grid spectroscopic map of noise measured near the impurity (the green square in **b**) at the same bias $V_{bias}$ = -2 mV. The ring feature is absent in noise. **d** Differential conductance spectra measured on and off the impurity, denoted by the red and blue crosses in **b**, respectively. **e,f** Noise spectrum (**e**) and its corresponding effective charge (**f**) spectrum taken at off impurity position, showing a step from 1$e$- to 2$e$-noise. Error bars in **e** are determined by the fluctuation of the current noise in time, yielding a standard deviation of 6.77 fA$^2$/Hz. Error bars in **f** correspond to numerical solutions of Eq. 2 in the main text, using values of upper and lower bounds indicated the error bars in **e**, respectively. Setup conditions: **a**, **b**, and **d**, $V_{set}$ = -8 mV, $I_{set}$ = 4 nA; **c**, $V_{set}$ = -2 mV, $I_{set}$ = 800 pA, $R_J$ = 2.5 MOhm; **e**, $R_J$ = 10 MOhm.

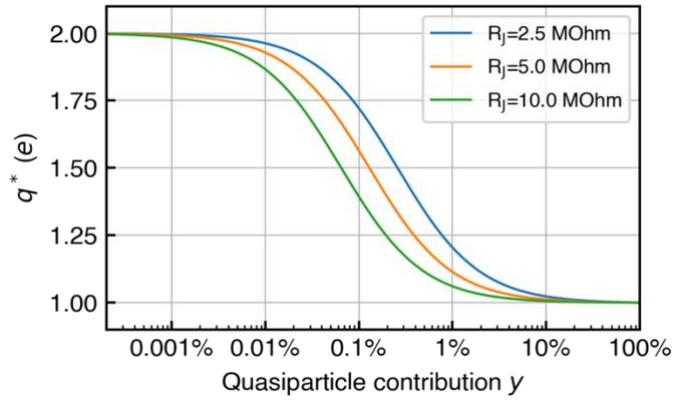

**Supplementary Figure 7. Simulation of effective charge as a function of quasiparticle contribution.** Three different junction resistance $R_J$ (hence transparency) of 2.5, 5, and 10 MOhm in our experiments are used as an input in the model (see Supplementary Note 5 for details of the model).

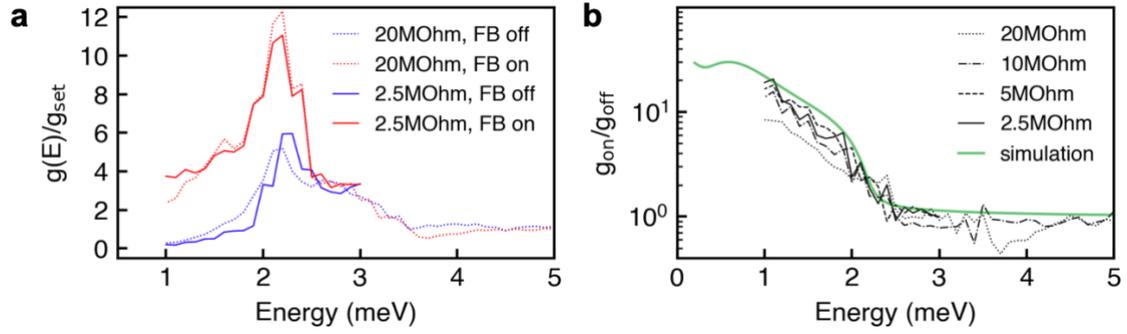

**Supplementary Figure 8. Transparency of the junction with and without feedback control. a** Differential conductance spectra, normalized by setup conductance $g_{set}$, for different junction resistance (solid line, $R_J = 2.5$ MOhm; dotted line, $R_J = 20$ MOhm). Red and blue lines correspond to spectra taken with (FB on) and without (FB off) feedback control, respectively. **b** Measured conductance ratio for different junction resistance. The green line shows the simulation result (see Supplementary Note 6), in good agreement with all the data. The small dip near zero energy originates from the difference between the energy gaps of the tip and sample. Note the logarithmic scale of the vertical axis.

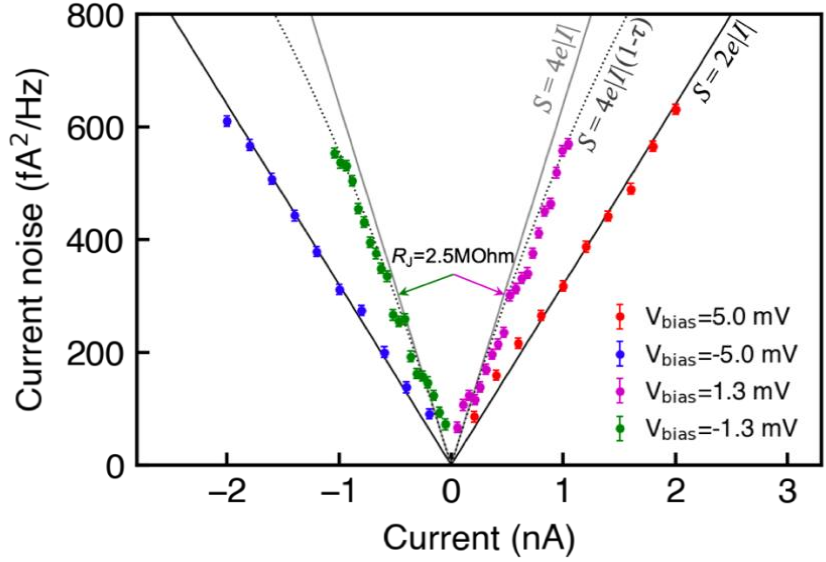

**Supplementary Figure 9. Linear dependence of current noise as a function of tunneling current in the high-bias limit** ($|eV_{bias}| >> k_B T = 0.2$ **meV for** $T = 2.3$ **K).** Red (blue) data show current noise measured on NbSe$_2$ at a fixed positive (negative) bias $eV_{bias} = \pm 5$ meV outside the gap. Purple (green) data show current noise measured at a fixed positive (negative) bias $eV_{bias} = \pm \Delta_t = \pm 1.3$ meV in the gap. Purple (green) arrows show the condition where measurements in Fig. 4 are taken. Error bars are determined by the fluctuation of the current noise in time, yielding a standard deviation of 9.25 fA$^2$/Hz. The lines show theoretical expectations of shot noise, for $q^* = 1e$ (black), $q^* = 2e$ (gray), and $q^* = 2e$ with correction for nonlinear junction transparency (black dotted line).

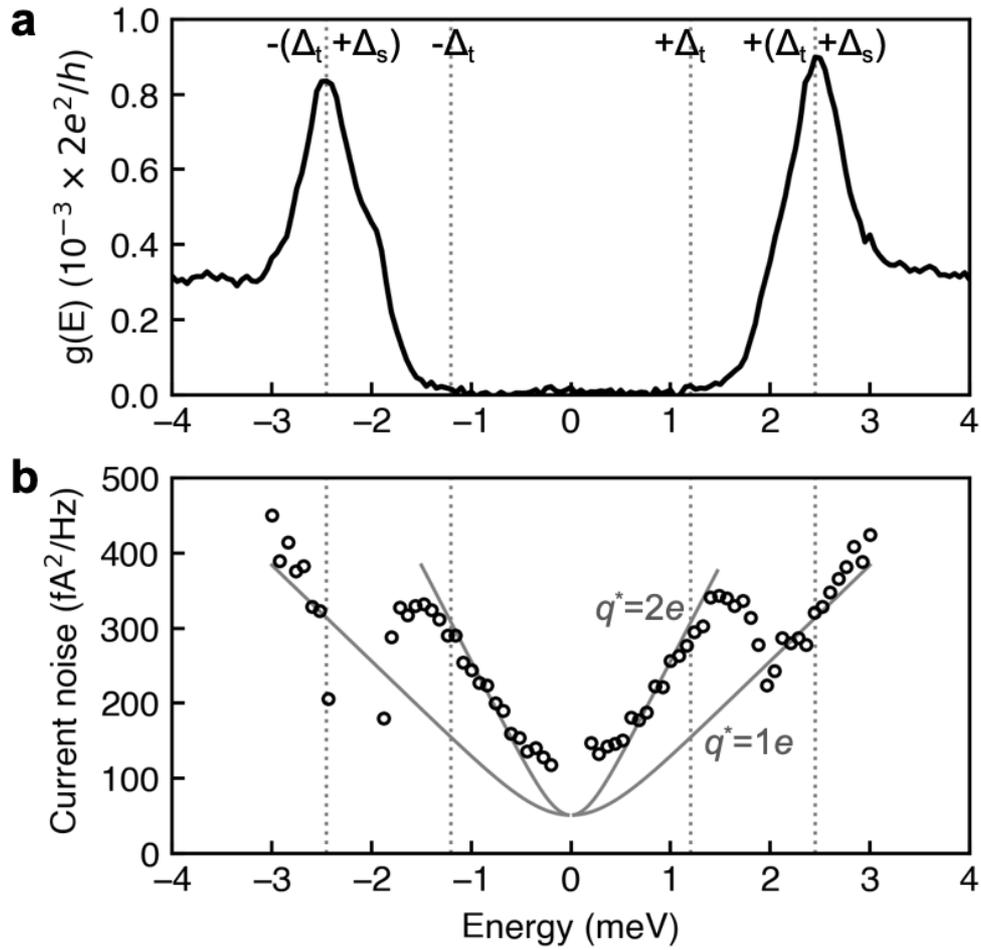

**Supplementary Figure 10. Differential conductance and noise spectra measured on NbSe₂ at zero field. a** Differential conductance spectrum and **b** corresponding noise spectrum measured at $B = 0$ T and a random location on a NbSe$_2$ sample. Noise data increase from $q^* = 1e$ curve starting at $\pm(\Delta_t + \Delta_s)$, towards to $q^* = 2e$ curve inside the gap. Setup conditions: **a**, $V_{set} = -5$ mV, $I_{set} = 200$ pA; **b**, $R_J = 2.5$ MOhm.